# Coherent phonon Rabi oscillations with a high frequency carbon nanotube phonon cavity

Dong Zhu,[1,2]† Xin-He Wang,[3,4]† Wei-Cheng Kong,[1,2]† Guang-Wei Deng,[1,2] Jiang-Tao Wang,[3,4] Hai-Ou Li,[1,2] Gang Cao,[1,2] Ming Xiao,[1,2] Kai-Li Jiang,[3,4] Xing-Can Dai,[3,4] Guang-Can Guo,[1,2] Franco Nori,[5,6] and Guo-Ping Guo[1,2]*

(September 20, 2016)

[1]Key Laboratory of Quantum Information, University of Science and Technology of China, Chinese Academy of Sciences, Hefei 230026, China.

[2]Synergetic Innovation Center of Quantum Information and Quantum Physics, University of Science and Technology of China, Hefei, Anhui 230026, China.

[3]State Key Laboratory of Low-Dimensional Quantum Physics, Department of Physics and Tsinghua-Foxconn Nanotechnology Research Center, Tsinghua University, Beijing 100084, China.

[4]Collaborative Innovation Center of Quantum Matter, Beijing 100084, China.

[5]CEMS, RIKEN, Wako-shi, Saitama 351-0198, Japan

[6]Physics Department, University of Michigan, Ann Arbor, Michigan 48109-1040, USA

†These authors contributed equally to this work.

*Correspondence to: gpguo@ustc.edu.cn

**Abstract**

Phonon-cavity electromechanics allows the manipulation of mechanical oscillations similar to photon-cavity systems. Many advances on this subject have been achieved in various materials. In addition, the coherent phonon transfer (phonon Rabi oscillations) between the phonon cavity mode and another oscillation mode has attracted many interest in nano-science. Here we demonstrate coherent phonon transfer in a carbon nanotube phonon-cavity system with two mechanical modes exhibiting strong dynamical coupling. The gate-tunable phonon oscillation modes are manipulated and detected by extending the red-detuned pump idea of photonic cavity electromechanics. The first- and second-order coherent phonon transfers are observed with Rabi frequencies 591 kHz and 125 kHz, respectively. The frequency quality factor product $fQ_m \sim 2 \times 10^{12}$ Hz achieved here is larger than $k_B T_{\text{base}}/h$, which may enable the future realization of Rabi oscillations in the quantum regime.

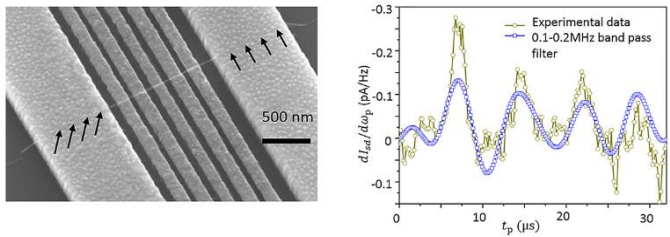

**Introduction**

Mode coupling of mechanical resonators has recently attracted considerable attention in nanotechnology for both practical applications and fundamental studies, such as mass sensing[1, 2], charge detection[3], band-pass filters[4], signal amplification[5], phonon-mediated electron interaction[6], chaotic dynamics[7] and nonlinearity[8-11].

To date, pioneering experiments have demonstrated classically coherent phonon manipulations in a single GaAs-based mechanical resonator[12] and two coupled GaAs-based mechanical beams[13] using dynamical coupling, and a SiN resonator[14] using a linear-coupling method different from the dynamical coupling used here. The manipulated phonon frequencies were of the order of 10 Hz, 100 kHz and several MHz, respectively. And the Rabi frequencies were typically around 40Hz, 100 Hz and 8 kHz, respectively. However, to further achieve coherent phonon manipulation in the quantum regime[15] ( $fQ_m > k_B T_{\text{base}}/h$ ), higher-frequency mechanical resonators are also desirable due to fewer-phonon occupation at a fixed temperature $T_{\text{base}}$. For high-frequency resonators (such as carbon nanotubes[16], $MoS_2$[17], and graphene[18] systems), the large-frequency dispersion and the natural multi-modes offer a platform to use the higher-frequency mode as a phononic cavity. Dynamical strong-mode couplings have been reported in previous works[16-18], which show the advantage of coupling two mechanical modes with arbitrary frequency differences. In spite of these remarkable achievements, there is still a lack of coherent phonon manipulation in these systems[16-18].

Carbon nanotube (CNT) mechanical resonators[19-26] are well-known not only for their high-quality factors[24] and low mass, but also for their high frequencies. Nonlinear strong mode-coupling in carbon nanotube resonators has been studied in a few works[9, 10], while dynamical strong mode-coupling has not been implemented until very recently[16]. The exceptional properties of CNTs mean that the phonon Rabi oscillations implemented by a CNT phonon cavity should exhibit a superior performance.

In this letter, we report the successful implementation of coherent phonon Rabi oscillations, with a ~ 50 MHz resonant mechanical frequency, in a CNT resonator. By combining previous ideas on the dynamical mode-coupling method in separate beams[17] and linear mode-coupling in a single resonator[14], we achieved dynamical strong mode-coupling between two distinct modes in a single CNT resonator. Analogous to photonic cavity-mechanical systems[27-31], the higher-frequency mode in our system can be treated as a phonon cavity [17, 18, 32], and the lower-frequency mode is shown to be parametrically-coupled to this microwave phonon cavity. Coherent phonon transfers are observed in this mechanical phonon cavity system with Rabi frequency 591 kHz (for the first-order) and 125 kHz (for the second-order), respectively. Furthermore, the achieved frequency quality factor product, $fQ_m \sim 2 \times 10^{12}$ Hz, fulfills the standard to reach the quantum regime ($fQ_m > k_B T_\text{base}/h$) after further cooling[15].

**Results**

We design and fabricate a carbon nanotube resonator suspended over the source and drain electrodes, which are 150 nm thick and 800 nm wide [Figure 1(a)]. The trench is designed to be 1200 nm wide and 100 nm deep; also five gates (with 50 nm thickness and 120 nm width) are located underneath the CNT. The CNT is grown by chemical vapor deposition and is typically double-walled and about 2-3 nm in diameter. The inner shell tube of the double-walled CNT is then drawn out using a micro-manipulation technique, which is then transferred and precisely positioned on the pre-designed metal electrodes [source (S) and drain (D)]. The scanning electron microscope (SEM) picture [Figure 1(b)] shows that the CNT is suspended quite well over the gate electrodes and the sample is very clean.

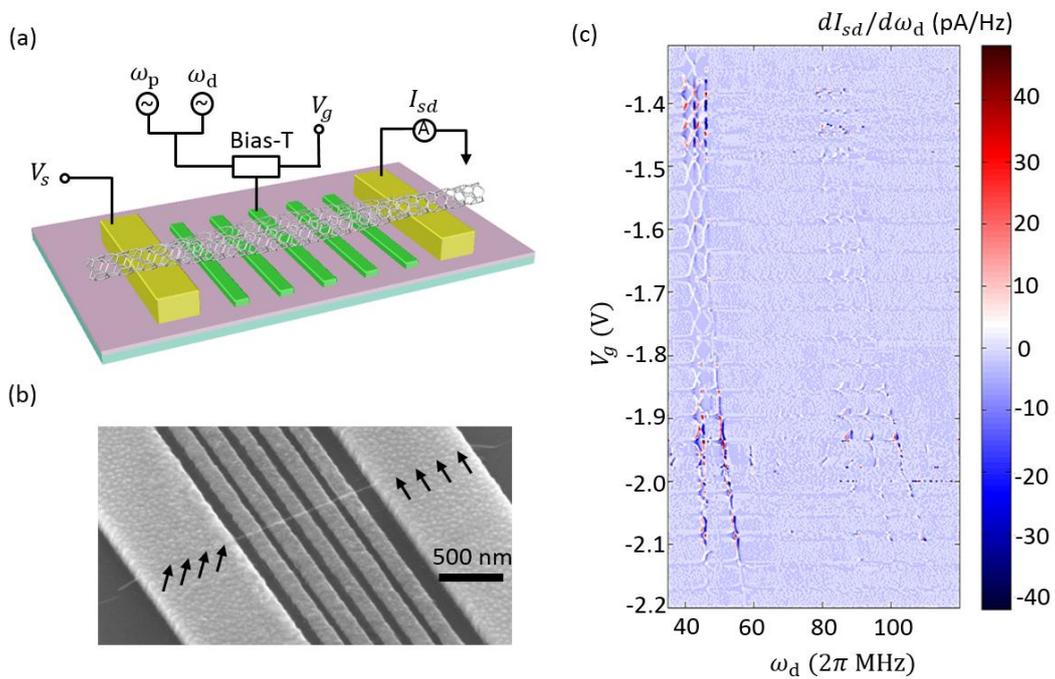

**Figure 1. Carbon nanotube resonator with large frequency tunability of various flexural modes.** (a) Schematic of the sample structure and the measurement circuit. A carbon nanotube (CNT) resonator is in contact with the source and drain electrodes (800 nm wide) and suspended over five gates (120 nm wide and spaced by 80 nm) with a trench of 1200 nm and a depth of 100 nm. (b) Scanning electron micrograph of the device, with a scale bar of 500 nm. (c) Large-frequency tunability of various flexural modes with the middle gate voltage under a drive microwave power $P_d = -43$ dBm.

Two distinct first mechanical modes (~ 50 MHz) and three second mechanical modes (~ 100 MHz) vary largely when changing the DC gate voltage. Here the derivative of the transport current with respect to the frequency of drive microwave $dI_{sd}/d\omega_d$ to remove the Coulomb blockade and clearly show all the possible mechanical modes (See the transport current data in Fig. S2 of the supporting informations III).

The measurements are performed in a dry refrigerator Triton system with base temperature of approximately 10 mK and a pressure typically under $10^{-6}$ torr. We use an electrical approach to actuate and detect the mechanical modes of the CNT. A biased voltage $V_s$ is added to the *S* electrode and the DC transport current $I_{sd}$ of the CNT is measured at the *D* electrode with a commercial multimeter. We use a bias-Tee to apply both DC and AC voltages on the third of the five gates, where the DC voltage is used to tune the chemical potential of the quantum dot formed on the local region of the CNT above the gate and the AC voltage is used to actuate the CNT resonator. In this experiment, two pump and drive microwave signals ($\omega_p$ and $\omega_d$) are combined on the AC port [Figure 1(a)]. When the frequency of drive microwave $\omega_d$ approaches the resonance frequency of the CNT, the DC transport current $I_{sd}$ shows a distinct peak as the signal of the mechanical vibration, which is due to the change of the displacement-modulated capacitance of the suspended CNT quantum dot (see the Supporting informations I). The various vibrational modes are shown in Figure 1(c) with an input drive signal power $P_d = -43$ dBm, where two distinct mechanical modes of the first-order vibration ($\omega_1$ for the lower frequency mode and $\omega_2$ for the higher mode) with a frequency of approximately 50 MHz (varying largely when changing the DC gate voltage, approximately 20 MHz/V) are so strongly coupled to a quantum dot that they cannot be distinguished for small gate voltages. This phenomenon is due to the high drive power. We can reasonably infer that: (1) the mode with lower resonant frequency mainly vibrates in-plane parallel to the gate electrode and (2) the higher mode vibrates out-of-the-plane because the electrostatic force between the gate and the CNT is stronger in the vertical direction, so the vertical vibration tension in the CNT is larger than the parallel vibration tension, causing a larger spring effect[10, 33]. Figure 1(c) also shows three weak mechanical modes of the second-order vibration with resonant frequencies of approximately 100 MHz. Because of the symmetric structure of our measurement scheme, the vibration modes with an odd number of wave nodes cannot apparently change the capacity between the gate and the CNT. The resonant frequencies for the first vibration modes (~50 MHz) and second vibration modes (~100 MHz) are approximately in agreement with the elastic string model[33], which has a simple relation on the multiple higher-vibration frequencies and first-vibration frequencies: $f_n = nf_1$.

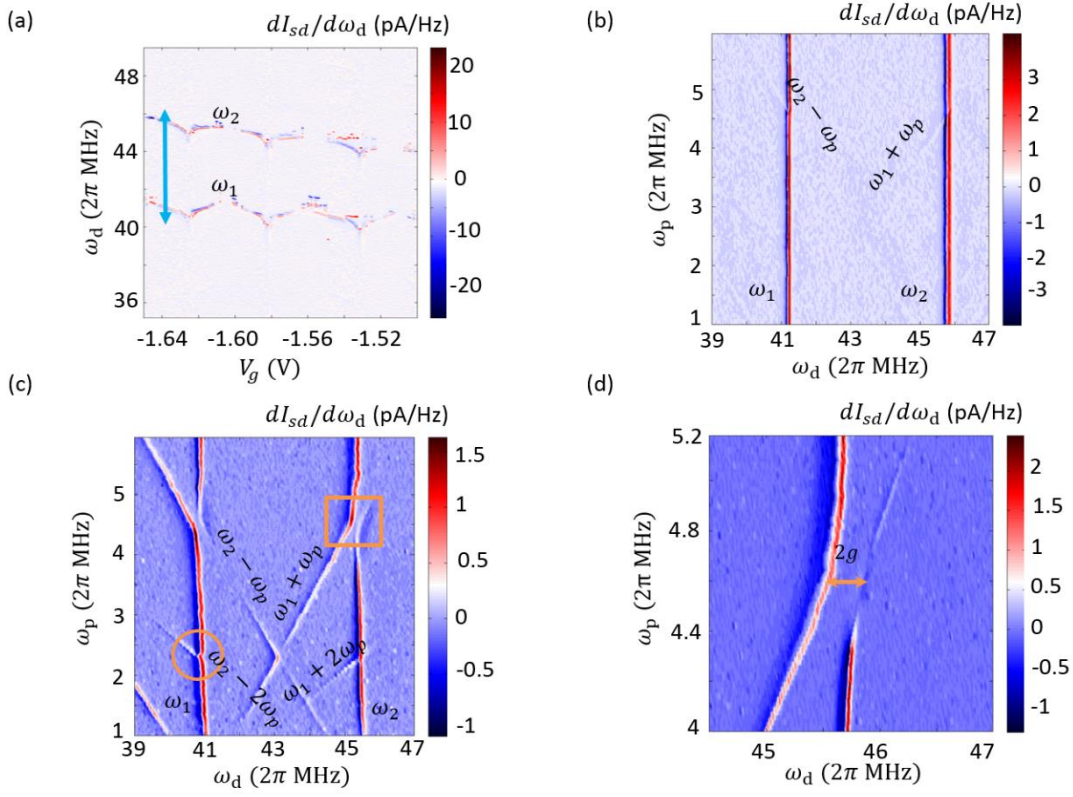

**Figure 2. Dynamical strong mode-coupling via normal-mode splitting.** (a) The two distinct first mechanical modes under a weak drive power of $P_d = -70$ dBm. The lower-frequency mode is denoted as "oscillation mode 1" and the higher-frequency mode is denoted as "phonon cavity mode 2". The normal-mode splitting is performed at the gate voltage $V_g = -1.64$ V (see the arrow). (b) The Stokes and anti-Stokes sidebands of the two modes with a pump power $P_p = -49$ dBm. Only the first-order sidebands are shown because of the weak pump power. These sidebands cause dynamical coupling between the oscillation mode and phonon cavity mode. (c) First-order normal-mode splitting with a higher pump power $P_p = -44$ dBm, when the pump frequency $\omega_p$ approaches $\Delta\omega$ (see the rectangle). Both the oscillation mode and the phonon cavity mode show strong dynamical mode-coupling. The second-order mode-coupling is not apparent at $\omega_p = \Delta\omega/2$ for the low-pump power (see the circle).

Both first- and second-order Stokes and anti-Stokes sidebands of the two mode are shown. (d) Larger normal mode splitting of the phonon cavity at $\omega_p = \Delta\omega$ with a pump power $P_p = -34$ dBm. The coupling rate is approximately $g = 2\pi \times 225$ kHz (see the arrow).

In this experiment, the two distinct mechanical modes of the first-order vibration are selected to implement the phonon Rabi oscillations or coherent phonon manipulations. Hereafter, we denote the lower-frequency mode as oscillation mode 1 and the higher-frequency mode as phonon cavity mode 2. To improve the quality factor of the resonator, the drive microwave power is further reduced to $P_d = -70$ dBm, and Figure 2(a) shows the resonant frequencies of modes 1 and 2 with different DC gate voltages measured by single-electron tunneling. Under weak-drive microwave power, modes 1 and 2 are distinctly separated from each other, with a frequency difference $\Delta\omega \sim 2\pi \times 5$ MHz. The quality factors for both modes are ~50,000, and the corresponding decoherence rate $\gamma \sim 2\pi \times 900$ Hz.

We further perform a strong dynamical coupling between oscillation mode 1 and the phonon cavity mode 2 by showing the normal-mode splitting, for which a red-detuned (relative to the phonon cavity mode) parametric pump microwave $\omega_p$ $(\omega_p \approx \omega_2 - \omega_1 = \Delta\omega)$ is applied to the CNT combined with the previous drive microwave $\omega_d$ with $P_d = -70$ dBm [see Figure 1(a)]. By applying a gate voltage $V_g = -1.64$ V, we tune the oscillation mode 1 frequency to $\omega_1 = 2\pi \times 41.12$ MHz and the resonant frequency of the phonon cavity mode to $\omega_2 =$ 45.68 MHz and $\Delta\omega = 2\pi \times 4.56$ MHz, which are shown by the arrow in Figure 2(a). Figure 2(b) shows the two modes while scanning the pump microwave frequency with a pump power $P_p = -49$ dBm. The Stokes and anti-Stokes sidebands of the two modes emerge and these sidebands cause dynamical coupling between the oscillation mode and phonon cavity mode. $dI_{sd}/d\omega_d$ is used in Fig. 2 (b)-(d) to reduce the noise and clearly show the coupling between the two mechanical modes. Here at different pump frequency $\omega_p$, we measure the transport current as scanning the drive frequency $\omega_d$ and the noise is produced along the axis of pump frequency $\omega_p$. Normal-mode splitting is demonstrated in Figure 2(c) for both modes with a slightly larger pump power $P_p = -44$ dBm, when the pump frequency $\omega_p$ approaches $\Delta\omega$. The pumped microwaves produce a series of sidebands of the two modes with phonon frequency $\omega_{1,2} \pm n\omega_p$. When the Stokes sideband of mode 2 matches the resonance frequency of mode 1, that is, $\hbar\omega_1 \sim (\hbar\omega_2 - \hbar\omega_p)$, or the anti-Stokes sideband of mode 1 matches the resonance frequency of mode 2, $\hbar\omega_2 \sim (\hbar\omega_1 + \hbar\omega_p)$, avoided crossing occurs. This avoided crossing signifies that dynamical coupling between the oscillation mode and the phonon cavity is in the strong-coupling regime. The coupling strength $g$ is about half the frequency splitting [Figure 2(d)] and can also be tuned by the amplitude of the pump microwave. Figure 2(d) shows the normal-mode splitting of the phonon cavity mode $\omega_2$ with a larger pump power $P_p = -34$ dBm, and the coupling rate is increased to be $g = 2\pi \times 225$ kHz. This is much larger than that in GaAs-based mechanical resonators[13, 32] (typically 100 Hz) and approximately 27 times larger than in silicon nitride devices[14] (8.3 kHz) which use a linear coupling method different to the dynamical coupling used here. The strong coupling between the oscillation mode and the phonon cavity can also be quantified by a figure of merit called cooperativity, defined as $C = 4g^2/(\gamma_1\gamma_2)$, which is ~250,000 in our system and *much higher* than that in graphene[18] ($C$ as high as 60), $MoS_2$ resonators[17] ($C = 2,066$), and GaAs-based mechanical resonators[13] ($C = 90$). The frequency quality factor product $fQ_m \sim 2 \times 10^{12}$ Hz in our device is also the *highest* compared with that reported in other similar studies[13, 17, 18, 32], and is very close to the standard of phonon-cavity mechanics in the quantum regime[15] at room temperature ($fQ_m > k_B T_{\text{room}}/h$).

The strong dynamical coupling between the oscillation mode $\omega_1$ and the phonon cavity mode $\omega_2$ offers a platform for coherent phonon transfer between a phonon cavity and an oscillator (phonon Rabi oscillation). Figure 3(a) shows the time-domain pulse sequence we used to achieve this goal. The detailed measurement scheme of the phonon Rabi oscillation is discussed in the supporting informations V. First, a drive microwave with frequency $\omega_d$, matching the phonon cavity resonant frequency $\omega_2$, is applied to actuate the phonon cavity mode for $\sim t_d$ and then turned off, which is realized by a frequency mixer and arbitrary waveform generator (AWG). At the time when the drive microwave is turned off, the red-detuned parametric pump microwave with $\omega_p = \omega_2 - \omega_1 = \Delta\omega$ is turned on for a time $t_p$, which causes the coherent energy exchange between the phonon cavity and the oscillation mode. This is the first-order coherent process ($n = 1$). The second-order coherent process ($n = 2$) occurs for the case when $\omega_p = \Delta\omega/2$, while the normal-mode splitting for this case is not apparent in Figure 2(c) (the circle).

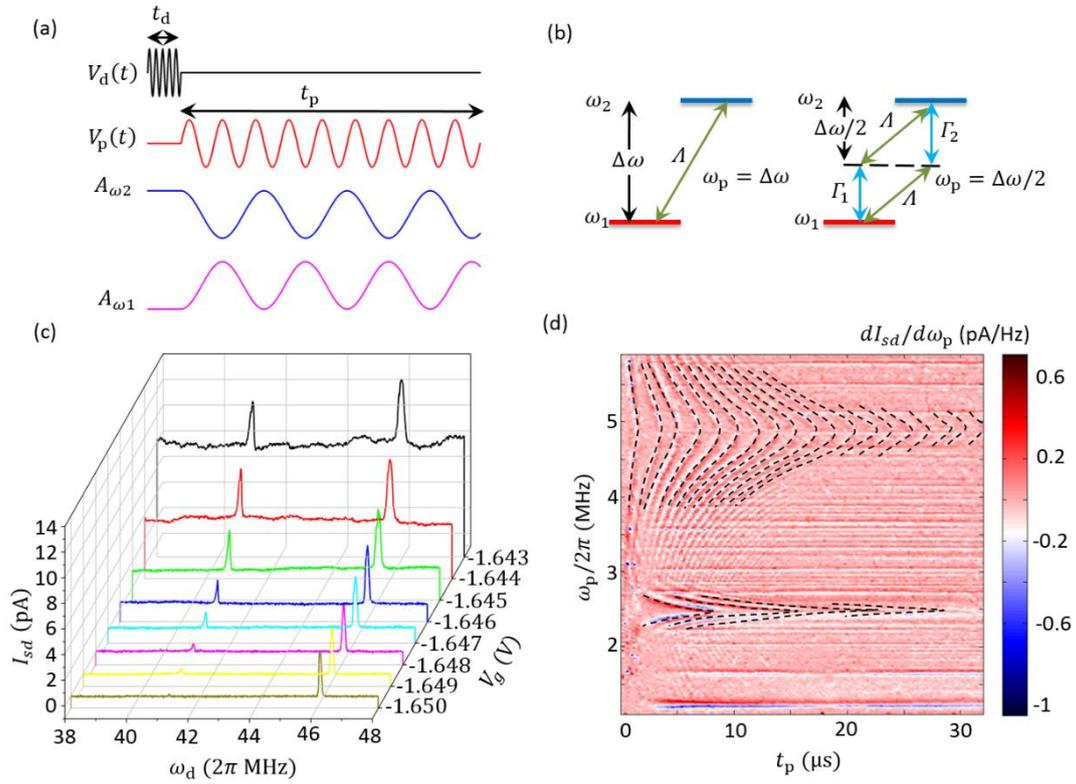

**Figure 3. Phonon Rabi oscillation between the oscillation mode and phonon cavity mode.** (a) The time-domain pulse sequence used to implement phonon Rabi oscillations. (b) Schematic of first- and second-order phonon Rabi oscillation processes. (c) DC transport current by scanning the drive microwave frequency with different gate voltages, from $-1.643$ V to $-1.650$ V. The voltage step is $-0.001$ V and drive power $P_d = -53$ dBm. (d) Phonon Rabi oscillations with a large pump power $P_p = 3$ dBm. 19 distinct cycles of coherent energy exchanges between the oscillation mode and the phonon cavity mode at $\omega_p = \Delta\omega = 4.9$ MHz (the first order) are shown and 4 cycles of second-order coherent energy exchange at $\omega_p = \Delta\omega/2 = 2.45$ MHz are also observed for the large pump power. $dI_{sd}/d\omega_p$ is used again to increase the signal-to-noise ratio. Here at different pump time $t_p$, we measure the transport current as scanning

the pump frequency $\omega_\mathrm{p}$ and the noise is produced along the axis of pump time $t_\mathrm{p}$. The original results is provided in Fig. S5 in the supporting informations VI.

For the single CNT resonator, we have to identify the contributions from each of the two modes to the DC transport current to manipulate the phonon Rabi oscillation. However, this is very difficult when both of the two modes exist at the same time, because the influence of mechanical vibrations on the transport current through single-electron tunneling is an average effect for high-frequency motion. To solve this problem, we introduce a useful and novel refinement in our experiment. By carefully changing the gate voltage, we find that the oscillation mode 1 can be tuned to where the transport current is not sensitive, so that only the phonon cavity mode 2 can be detected, which is further used to perform the phonon Rabi oscillations (See the Supporting informations VII). Figure 3(c) shows the DC transport current when scanning the drive microwave frequency for different gate voltages: from $-1.643$ V to $-1.650$ V, in steps of $-0.001$ V. Note that when the gate voltage is smaller than $-1.646$ V, the oscillation mode with lower resonant frequency cannot be sensitively detected by the DC transport current, while the phonon cavity mode with a higher resonant frequency is not affected. When gate voltage is $-1.649$ V or $-1.650$ V, the oscillation mode is almost entirely not detected by the current, which enables us to only detect the phonon cavity mode when there is an energy exchange between these two modes caused by the pump. By applying a gate voltage $V_g = -1.650$ V, we tune the oscillation mode 1 frequency to $\omega_1 = 2\pi \times 41.20$ MHz and the resonant frequency of the phonon cavity mode to $\omega_2 = 46.10$ MHz and $\Delta\omega = 2\pi \times 4.9$ MHz. Figure 3(d) shows the phonon Rabi oscillations with pump power $P_\text{p} = 3$ dBm. The drive power is $P_\text{d} = -53$ dBm and the drive frequency matches the phonon cavity mode $\omega_2 = 46.10$ MHz. The higher drive power allows the phonon cavity mode with a larger amplitude which will help to increase the electric signal strength of the Rabi oscillations, while the quality factor will decrease because of the stronger coupling between single-electron tunneling and nanomechanical motion[21]. The DC transport current dependence at $\omega_\text{p} = \Delta\omega = 4.9$ MHz shows 19 distinct cycles of energy exchanges between the oscillation mode and the phonon cavity mode which is better than the results (8 cycles) obtained from two coupled GaAs-based mechanical resonators[13]. The second-order process at $\omega_\text{p} = \Delta\omega/2 = 2.45$ MHz with 4 cycles of energy exchanges is shown in Figure 3 (d), and is a bit weaker than the first-order ones.

To understand the dynamics of the phonon Rabi oscillation, we model the dynamics of our system using two coupled vibration modes with a time-varying spring constant. The Hamiltonian[32] is:

$$H = \frac{p_c^2}{2m_\text{eff}} + \frac{1}{2}m_\text{eff}\left(\omega_2^2 + \Gamma_2 V_p^2\cos(\omega_p t)\right)q_c^2 + \frac{p_o^2}{2m_\text{eff}}$$
$$+ \frac{1}{2}m_\text{eff}\left(\omega_1^2 + \Gamma_1 V_p^2\cos(\omega_p t)\right)q_o^2 + m_\text{eff}\Lambda V_p^2\cos(\omega_p t)q_c q_o$$

where $q_o, p_o$ represent the oscillation mode 1 and $q_c, p_c$ represent the phonon cavity mode 2. Also, $\Lambda$ is the inter-modal coupling coefficient and $\Gamma_{1,2}$ are the intra-modal coupling coefficients for mode 1 and mode 2, respectively. The effective mass of the two modes are not distinguished as discussed in the supporting informations II. Figure 3(b) shows the schematics of the first- and second-order phonon Rabi oscillation processes. For the first-order coherent process, the inter-modal coupling term $\Lambda V_p^2$ is responsible for the energy exchange, with the pump frequency matching the frequency difference of the two modes $\omega_p = \Delta\omega$. Additionally, the coupling rate[13, 17, 18] is given by $g \sim \frac{\Lambda V_p^2}{2\sqrt{\omega_1 \omega_2}}$. For the second coherent process, the inter-modal coupling term $\Lambda V_p^2$ combined with the intra-modal coupling term $\Gamma_{1,2} V_p^2$ is responsible for the higher-order energy exchange between the modes [Figure 3(b)].

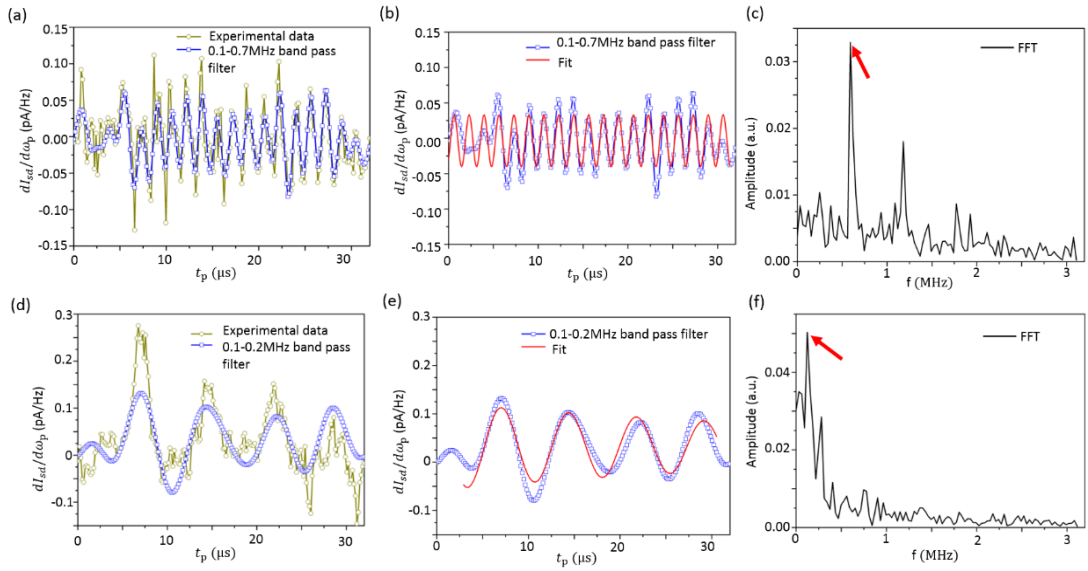

**Figure 4. Fitting the time domain data for the phonon Rabi coupling rate.** (a) First-order phonon Rabi oscillations at $\omega_p = \Delta\omega = 4.9 \text{ MHz}$ from Figure 3(d). Data in dark yellow are the experimental data and the data in blue are under a $0.1 \text{ MHz} - 0.7 \text{ MHz}$ band pass filter of the experimental data, which are in good agreement with the guide lines shown in Figure 3(d). (b) The first-order coupling rate is as high as $g_1 = 2\pi \times 591 \text{ kHz}$ by fitting the filtered data. (c) The fast Fourier transform (FFT) of the experimental data shows a distinct peak at $591 \text{ kHz}$ (see the arrow). (d) Second-order phonon Rabi oscillations at $\omega_p = \Delta\omega/2 = 2.45 \text{ MHz}$ from Figure 3(d). Data in dark yellow are the experimental data and the data in blue are under a $0.1 \text{ MHz} -$

0.2 MHz band pass filter of the experimental data, which is in good agreement with the guide lines in Figure 3(d). (e) Second-order coupling rate is $g_2 = 2\pi \times 125$ kHz by fitting the filtered data. (f) The FFT of the experimental data shows a distinct peak at 125 kHz (see the arrow).

The coupling rate, or the Rabi frequency, $g$ can be determined by fitting the time-domain data of the energy-exchange rate at $\omega_p = \Delta\omega$ and $\omega_p = \Delta\omega/2$, respectively. Figures 4(a) and (d) show the first-order and second-order coherent phonon Rabi oscillations, respectively, between the oscillation mode 1 and the phonon cavity mode 2. For the first-order oscillation, a $0.1\text{ MHz} - 0.7\text{MHz}$ band pass filter is used to remove the noise signal of the experimental data. As a result, the filtered signal in Figure 4(a) is in good agreement with the guide lines in Figure 3(d). The first-order coupling rate is $g_1 = 2\pi \times 591$ kHz by fitting the filtered data [Figure 4(b)], which agrees with the FFT of the original experimental data [see the arrow in Figure 4(c)]. Similarly, the second-order coupling rate is $g_2 = 2\pi \times 125$ kHz by fitting the filtered data with a $0.1\text{ MHz} - 0.2\text{ MHz}$ band pass filter [Figure 4(e)], which also agrees with the FFT of the original experimental data [see the arrow in Figure 4(f)]. The coupling rate is higher than the previous normal-mode splitting experiment [Figure 2(c)] because of the large pump power used here. Compared again with similar studies for GaAs-based mechanical resonators[13] and silicon nitride devices[14], our result is ~5,900 times larger than that for GaAs-based mechanical resonators (typically ~100 Hz) and about 70 times larger than that for silicon nitride devices (8.3 kHz).

**Conclusion**

We have demonstrated coherent phonon transfer between two mechanical modes of a carbon nanotube resonator. Similar to photonic cavity electro-mechanics[27-32], the red-detuned pump technique is used to demonstrate the normal-mode splitting of the oscillation mode (low-frequency mode) and the phonon cavity (high-frequency mode). An electromagnetic pulse[13, 14] is applied to implement the coherent phonon Rabi oscillations in this phonon-cavity mechanical system. Compared with previous studies[12-14, 17, 18, 32], the present study shows higher-frequency classical phonon Rabi oscillations, a larger cooperativity $C$, and a higher-frequency quality-factor product[15] ( $fQ_m > k_B T_{\text{base}}/h$ ). The methods demonstrated here can be applied to other mechanical systems such as MoS$_2$ or graphene resonators[17, 18] and will also be highly valuable for future studies of quantum phonon cavity systems and electron-phonon coupling[20, 21, 23, 34]

**Acknowledgements**

This work was supported by the National Key R&D Program (Grant No.2016YFA0301700), the Strategic Priority Research Program of the CAS (Grant No. XDB01030000), the National Natural Science Foundation (Grant Nos. 11304301, 11575172, 61306150, and 91421303), and the Fundamental Research Fund for the Central Universities. It was also partially supported by a Grant-in-Aid for Scientific Research (A), and a grant from the John Templeton Foundation.

# References

1. Spletzer, M.; Raman, A.; Wu, A. Q.; Xu, X. F.; Reifenberger, R. *Applied Physics Letters* **2006,** 88, (25), 254102.
2. Gil-Santos, E.; Ramos, D.; Jana, A.; Calleja, M.; Raman, A.; Tamayo, J. *Nano letters* **2009,** 9, (12), 4122-7.
3. Okamoto, H.; Kitajima, N.; Onomitsu, K.; Kometani, R.; Warisawa, S.; Ishihara, S.; Yamaguchi, H. *Applied Physics Letters* **2011,** 98, (1), 014103.
4. Bannon, F. D.; Clark, J. R.; Nguyen, C. T.-C. *IEEE JOURNAL OF SOLID-STATE CIRCUITS* **2000,** 35, (4), 512.
5. Karabalin, R. B.; Lifshitz, R.; Cross, M. C.; Matheny, M. H.; Masmanidis, S. C.; Roukes, M. L. *Phys Rev Lett* **2011,** 106, (9), 094102.
6. Deng, G. W.; Zhu, D.; Wang, X. H.; Zou, C. L.; Wang, J. T.; Li, H. O.; Cao, G.; Liu, D.; Li, Y.; Xiao, M.; Guo, G. C.; Jiang, K. L.; Dai, X. C.; Guo, G. P. *Nano letters* **2016,** 16, (9), 5456-62.
7. Karabalin, R. B.; Cross, M. C.; Roukes, M. L. *Physical Review B* **2009,** 79, (16).
8. Westra, H. J.; Poot, M.; van der Zant, H. S.; Venstra, W. J. *Phys Rev Lett* **2010,** 105, (11), 117205.
9. Castellanos-Gomez, A.; Meerwaldt, H. B.; Venstra, W. J.; van der Zant, H. S. J.; Steele, G. A. *Physical Review B* **2012,** 86, (4).
10. Eichler, A.; del Alamo Ruiz, M.; Plaza, J. A.; Bachtold, A. *Phys Rev Lett* **2012,** 109, (2), 025503.
11. Eriksson, A. M.; Midtvedt, D.; Croy, A.; Isacsson, A. *Nanotechnology* **2013,** 24, (39), 395702.
12. Mahboob, I.; Nier, V.; Nishiguchi, K.; Fujiwara, A.; Yamaguchi, H. *Applied Physics Letters* **2013,** 103, (15), 153105.
13. Okamoto, H.; Gourgout, A.; Chang, C.-Y.; Onomitsu, K.; Mahboob, I.; Chang, E. Y.; Yamaguchi, H. *Nature Physics* **2013,** 9, (8), 480-484.
14. Faust, T.; Rieger, J.; Seitner, M. J.; Kotthaus, J. P.; Weig, E. M. *Nature Physics* **2013,** 9, (8), 485-488.
15. Norte, R. A.; Moura, J. P.; Groblacher, S. *Phys Rev Lett* **2016,** 116, (14), 147202.
16. Li, S. X.; Zhu, D.; Wang, X. H.; Wang, J. T.; Deng, G. W.; Li, H. O.; Cao, G.; Xiao, M.; Guo, G. C.; Jiang, K. L.; Dai, X. C.; Guo, G. P. *Nanoscale* **2016,** 8, (31), 14809-13.
17. Liu, C. H.; Kim, I. S.; Lauhon, L. J. *Nano letters* **2015,** 15, (10), 6727-31.
18. Mathew, J. P.; Patel, R. N.; Borah, A.; Vijay, R.; Deshmukh, M. M. *Nature nanotechnology* **2016,** 11, (9), 747-51.
19. Sazonova, V.; Yaish, Y.; Ustunel, H.; Roundy, D.; Arias, T. A.; McEuen, P. L. *Nature* **2004,** 431, (7006), 284-7.
20. Lassagne, B.; Tarakanov, Y.; Kinaret, J.; Garcia-Sanchez, D.; Bachtold, A. *Science* **2009,** 325, (5944), 1107-10.
21. Steele, G. A.; Huttel, A. K.; Witkamp, B.; Poot, M.; Meerwaldt, H. B.; Kouwenhoven, L. P.; van der Zant, H. S. *Science* **2009,** 325, (5944), 1103-7.
22. Eichler, A.; Moser, J.; Chaste, J.; Zdrojek, M.; Wilson-Rae, I.; Bachtold, A. *Nature nanotechnology* **2011,** 6, (6), 339-42.
23. Benyamini, A.; Hamo, A.; Kusminskiy, S. V.; von Oppen, F.; Ilani, S. *Nature Physics* **2014,** 10, (2), 151-156.
24. Moser, J.; Eichler, A.; Guttinger, J.; Dykman, M. I.; Bachtold, A. *Nature nanotechnology* **2014,** 9, (12), 1007-11.

25. Moser, J.; Guttinger, J.; Eichler, A.; Esplandiu, M. J.; Liu, D. E.; Dykman, M. I.; Bachtold, A. *Nature nanotechnology* **2013,** 8, (7), 493-6.
26. Schneider, B. H.; Singh, V.; Venstra, W. J.; Meerwaldt, H. B.; Steele, G. A. *Nature communications* **2014,** 5, 5819.
27. Teufel, J. D.; Donner, T.; Li, D.; Harlow, J. W.; Allman, M. S.; Cicak, K.; Sirois, A. J.; Whittaker, J. D.; Lehnert, K. W.; Simmonds, R. W. *Nature* **2011,** 475, (7356), 359-63.
28. Teufel, J. D.; Li, D.; Allman, M. S.; Cicak, K.; Sirois, A. J.; Whittaker, J. D.; Simmonds, R. W. *Nature* **2011,** 471, (7337), 204-8.
29. Palomaki, T. A.; Harlow, J. W.; Teufel, J. D.; Simmonds, R. W.; Lehnert, K. W. *Nature* **2013,** 495, (7440), 210-4.
30. Singh, V.; Bosman, S. J.; Schneider, B. H.; Blanter, Y. M.; Castellanos-Gomez, A.; Steele, G. A. *Nature nanotechnology* **2014,** 9, (10), 820-4.
31. Weber, P.; Guttinger, J.; Tsioutsios, I.; Chang, D. E.; Bachtold, A. *Nano letters* **2014,** 14, (5), 2854-60.
32. Mahboob, I.; Nishiguchi, K.; Okamoto, H.; Yamaguchi, H. *Nature Physics* **2012,** 8, (5), 387-392.
33. Ustunel, H.; Roundy, D.; Arias, T. A. *Nano letters* **2005,** 5, (3), 523-6.
34. Li, P. B.; Xiang, Z. L.; Rabl, P.; Nori, F. *Phys Rev Lett* **2016,** 117, (1), 015502.

# Supporting information for "Coherent phonon Rabi oscillations with a high frequency carbon nanotube phonon cavity"

Dong Zhu,[1,2]† Xin-He Wang,[3,4]† Wei-Cheng Kong,[1,2]† Guang-Wei Deng,[1,2] Jiang-Tao Wang,[3,4] Hai-Ou Li,[1,2] Gang Cao,[1,2] Ming Xiao,[1,2] Kai-Li Jiang,[3,4] Xing-Can Dai,[3,4] Guang-Can Guo,[1,2] Franco Nori,[5,6] and Guo-Ping Guo[1,2]*

[1]Key Laboratory of Quantum Information, University of Science and Technology of China, Chinese Academy of Sciences, Hefei 230026, China.

[2]Synergetic Innovation Center of Quantum Information and Quantum Physics, University of Science and Technology of China, Hefei, Anhui 230026, China.

[3]State Key Laboratory of Low-Dimensional Quantum Physics, Department of Physics and Tsinghua-Foxconn Nanotechnology Research Center, Tsinghua University, Beijing 100084, China.

[4]Collaborative Innovation Center of Quantum Matter, Beijing 100084, China.

[5]CEMS, RIKEN, Wako-shi, Saitama 351-0198, Japan

[6]Physics Department, University of Michigan, Ann Arbor, Michigan 48109-1040, USA

†These authors contributed equally to this work.

*Correspondence to: gpguo@ustc.edu.cn

## I. The actuation and detection of CNT mechanical motion using quantum dot

We use an electrical approach to actuate and detect the mechanical modes of the CNT[1]. The suspended CNT simultaneously serves as a quantum dot and a mechanical resonator. The DC voltage $V_g$ and AC voltage $\delta V_g$ on the middle gate of our sample induce the average charges $\langle q_{dot}\rangle = C_g(V_g + \delta V_g)$ on the quantum dot formed on the local region of the CNT above the gate, where $C_g$ is the capacitance between the gate and the CNT. Because the electrode structures on the two sides of the CNT are asymmetric, the distribution of charges on the gate electrode is inhomogeneous. According to the electrostatics, except the surface charges of the electrodes, there are also many charges distributing at the terminals, which break the symmetry of the in-plane electric potential (Fig. S1). So there should be two kinds of mechanical modes, the in-plane mode $u_x$ with frequency of $\omega_1$ and out-of-plane mode $u_z$ with frequency of $\omega_2$. The electrostatic force of the CNT can be expressed as,

$$F_i = -\frac{\partial \langle U\rangle}{\partial u_i} = \frac{1}{2}\frac{\partial C_g}{\partial u_i}\left(V_g + \delta V_g\right)^2 \tag{S1}$$

Here, $\frac{\partial C_g}{\partial u_i}$ is the derivative of the gate capacitance with respect to the displacement of CNT resonator.

The DC voltage of $V_g$ can deform the CNT both in the in-plane and out-of-plane directions by the static force $F_i^{DC} = \frac{1}{2}\frac{\partial C_g}{\partial u_i}V_g^2$, and the induced additional tensions change the frequencies of the mechanical resonances. We can infer that the resonance frequency of in-plane mechanical mode $\omega_1$ is lower than that of the out-of-plane mechanical mode $\omega_2$, because the in-plane static force $F_x^{DC}$ is weaker than the out-of-plane static force $F_z^{DC}$, causing the lifted degeneracy of the spring constant of the in-plane and out-of-plane modes.

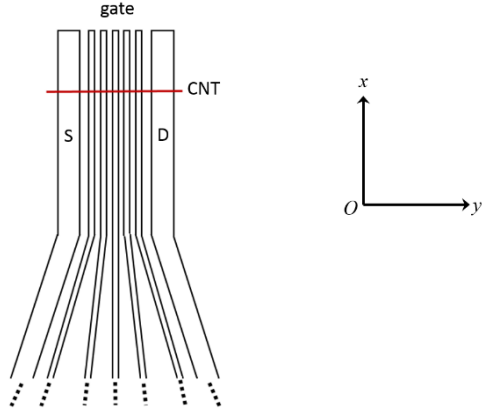

**Figure S1. The asymmetric structure of the sample electrodes.** The CNT is close to the terminals of the gate electrodes, breaking the symmetry of the in-plane electric potential.

The CNT resonator is actuated by the AC voltage of $\delta V_g(t) = \delta V_g^{\rm RF}\cos(\omega_{\rm d} t)$. When the drive frequency $\omega_{\rm d}$ approaches the resonance frequency $\omega_i$ of one mechanical mode, the periodic AC force $F_i^{\rm AC}(t) = \frac{\partial C_g}{\partial u_i} V_g \delta V_g(t)$ will effectively actuate the mechanical vibrations. The time-dependent displacement is $u_i(\omega_{\rm d},t) = A(\omega_{\rm d})\cos(\omega_{\rm d} t + \varphi)$, with driving frequency $\omega_{\rm d}$, phase factor $\varphi$, and amplitude of the mechanical oscillator

$$A(\omega_{\rm d}) = \frac{1}{m_{\rm eff}} \frac{\partial C_g}{\partial u_i} V_g \delta V_g^{\rm RF} \frac{1}{\sqrt{\left(\omega_i^2 - \omega_{\rm d}^2 + \frac{3}{4 m_{\rm eff}}\alpha A(\omega_{\rm d})^2\right)^2 + \frac{\omega_i^2 \omega_{\rm d}^2}{Q_i^2}}} \tag{S2}$$

Here, $\alpha$ is the nonlinear Duffing term, $m_{\rm eff}$ is the effective mass and $Q_i$ is the quality factor. The effective mass $m_{\rm eff}$ of the both in-plane and out-of-plane mechanical modes are same, which will be discussed later. The displacement-modulated capacitor of the suspended CNT can modify the transport current of the quantum dot, which is equivalent to an effective gate voltage as $V_{\rm eff}(\omega_{\rm d}, t) =$

$\frac{V_g}{C_g}\frac{\partial C_g}{\partial u_i}u_i(\omega_d, t)$. Therefore, the drain-source current changes with time as $I_{sd}(t) = \sum_n \frac{1}{n!}\frac{d^n I_{sd}^{DC}(V_g)}{dV_g^n}[V_{\text{eff}}(\omega_d, t)]^n$, in which $I_{sd}^{DC}(V_g)$ is the DC transport current of the quantum dot without the drive microwave, that is, the Coulomb blockade. The measured change of the DC transport current due to the mechanical vibration is

$$\Delta I_{sd} \approx \frac{1}{4}\frac{d^2 I_{sd}^{DC}(V_g)}{dV_g^2}\left[\frac{V_g}{C_g}\frac{\partial C_g}{\partial u_i}A(\omega_d)\right]^2 \tag{S3}$$

to the second order. When the drive frequency $\omega_d$ approaches the resonance frequency $\omega_i$ of one mechanical mode, the DC transport current $I_{sd}$ shows a distinct peak with the value of $\Delta I_{sd}$. The mechanical modes are detected by this signal.

**II. The effective mass of the in-plane and out-of-plane mechanical modes**

The effective mass $m_{\text{eff}}$ is determined by the conservation law of energy. The energy of the resonator can be written as,

$$E = \int_{-L/2}^{L/2} dy \left(\frac{1}{2}\rho\left(\frac{du_i(y,t)}{dt}\right)^2 + \frac{1}{2}\rho\omega_i^2 u_i(y,t)^2\right) = \frac{1}{2}m_{\text{eff}}\left(\frac{du_i(t)}{dt}\right)^2 + \frac{1}{2}m_{\text{eff}}\omega_i^2 u_i(t)^2$$

(S4)

The CNT is suspended between points of $y = -L/2$ and $y = L/2$. $\rho$ is the mass per unit length. $u_i(y,t) = u_i(t)f(z)$ denotes the displacement of the in-plane mode $u_x(y,t)$ or out-of-plane mode $u_z(y,t)$. $f(z)$ is waveform of the vibration modes, and we choose the $f(z) = \cos(\pi y/L)$ for the first-order vibration modes. $u_i(t)$ is the displacement at the center of the CNT, which denotes the displacement of the resonator. We find that $m_{\text{eff}} = \rho L/2$, which means that both the effective mass of the in-plane and out-of-plane mechanical modes are half of the total mass of the CNT.

**III. The Coulomb blockade and the resonance modes of CNT**

In Figure 1 (c) of the main paper, the derivative of the transport current with respect to the drive microwave frequency $dI_{sd}/d\omega_d$ is used to show the mechanical modes. In the experiment, the transport current is the measured physical quantity. $dI_{sd}/d\omega_d$ can remove the feature of Coulomb blockade and clearly show all the possible mechanical modes, which is very helpful.

Figure S2 (a) shows the original data of the Coulomb blockade of CNT quantum dot and the mechanical modes under the strong drive microwave power of $P_d = -43$ dBm. The Coulomb blockade line can be obtained by cutting the figure at a place far away from the resonance frequencies [Figure S2 (b)]. In these areas, the contribution to transport current by the mechanical vibrations can be neglected according to equation (S3).

When the middle gate voltage $|V_g|$ is larger than 1.9 V, the two mechanical modes of the first-order vibration mode (~ 50 MHz) are well separated and demonstrate the usual coupling feature with quantum dot, that is, the resonance frequency will emerge a dip when the gate voltage goes across one Coulomb peak. These resonance frequency dips were first reported in 2009[2, 3]. It is due to the strong phonon−electron tunneling interaction. The fluctuation of electron charges on the CNT induces the back-action force on the mechanical modes, softening the phonon modes[2, 3]. At the bottom of a Coulomb peak, the resonance signal of the mechanical modes of the first-order vibration is very good while the second-order vibration mode is very weak here as shown in Figure S3 (c), which is denoted by the green line in Figure S3 (a) and the green dot in Figure S3 (b) with a gate voltage $V_g = -2.068$ V.

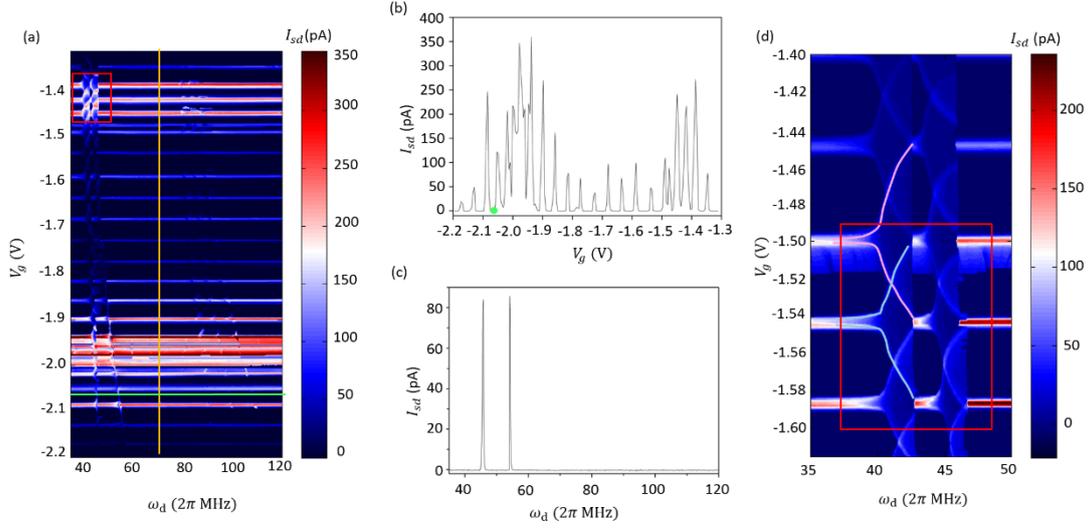

**Figure S2. Coulomb blockade of CNT quantum dot and the distinct two mechanical modes of the first-order vibrations.** (a) The mechanical modes showed by the DC transport current $I_{sd}$ with changing drive microwave frequency $\omega_d$ and the middle gate voltage $V_g$ under a strong drive microwave power $P_d = -43$ dBm. (b) The Coulomb blockade line of CNT quantum dot denoted by the vertical orange line in (a). (c) The distinct two mechanical modes of the first-order vibrations with good signal-to-noise ratio and a line width of ~ 100 kHz at the bottom of a Coulomb peak denoted by the horizontal green line in (a) and the green dot in (b) with $V_g = -2.068$ V. (d) The new detailed features of the coupling between mechanical modes and quantum dot at small gate voltages. The data was measured several days later than (a) and the three Coulomb peaks shifted which were denoted by the two red rectangles in (a) and (d).

When the gate voltage $|V_g|$ is smaller than 1.5 V, however, the two mechanical modes of the first-order vibration mode (~ 50 MHz) are so strongly coupled to the quantum dot that they cannot be distinguished with each other [Figure S2(d)]. Also, each mode shows strange coupling patterns as the gate voltage goes across the Coulomb peaks. This is a new phenomenon as shown in Figure S2 (d) caused by both the large drive microwave power and the small gate voltages. The strange patterns can be understand as the broadening of the resonance frequency dips. As shown in Figure S2 (d), the width of the two adjacent dips (the pink line and the light blue line) is broaden to 2 times of the gate voltage spacing of the Coulomb peaks. Because of the small gate voltage, the spacing of the two mechanical modes is so small as to make the coupling patterns cohering to each other. When the drive microwave power decreases, the strange coupling patterns disappear, as shown in Figure 2 in the main paper.

**IV. The dependence of the coupling rate on the pump power**

In the experiment, we perform a strong dynamical coupling between oscillation mode 1 and the phonon cavity mode 2 by showing the normal-mode splitting [see Figure 2 (c)-(d)]. When the pump microwave $P_\mathrm{p}$ is -49 dBm, the Stokes and anti-Stokes sidebands of the two modes emerge and are too weak to show the coupling between these two modes [Figure 2 (b)]. As the increase of the pump microwave power $P_\mathrm{p}$ from -44 dBm to -34 dBm, the Stokes and anti-Stokes sidebands of the two modes become strong and the coupling rate $g$ is tuned from $2\pi \times 165$ kHz to $2\pi \times 225$ kHz.

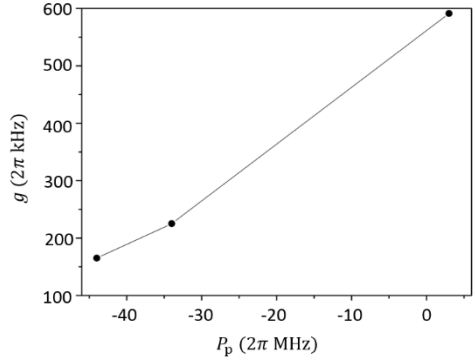

**Figure S3. The dependence of the coupling rate on the pump power**

Later, the phonon Rabi oscillations are demostrated at a pump power of 3 dBm, the coupling rate is extracted to be $2\pi \times 591$ kHz. Figure S3 shows the dependence of the coupling rate on the pump power. As the increase of the pump power over a large span, the coupling rate is slightly off the linear proportion to the pump power.

## V. The measurement scheme of the phonon Rabi oscillation

The demonstration of the strong coupling between the scillation mode 1 and the phonon cavity mode 2 are using the electrical measurement scheme shown in Figure 1 (a). The strong dynamical coupling offers a platform for coherent phonon transfer between a phonon cavity and an oscillator (phonon Rabi oscillation). Figure S4 shows the measurement scheme, which is used to produce the time-domain pulse sequence shown in Figure 3 (a).

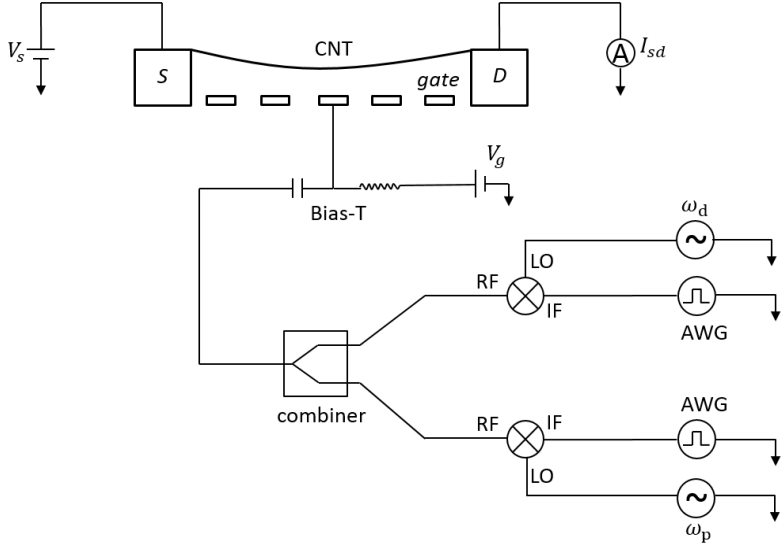

**Figure S4. The electrical measurement scheme of the phonon Rabi oscillation**

The two microwave tones (drive and pump tones) are mixed up with two control pulses respectively. The control pulse sequences are designed and generated by two synchronized AWG (Tektronix AWG7082C). Then they are combined and added to the same center back gate. The drive tone and the pump tone strongly couple the oscillation mode 1 and the phonon cavity mode 2, only in this time the two microwave tones are added at separate time intevals. The drive tone $\omega_d$ is turned on for a given time $t_d$ controlled by the length of the corresponding control pulse, and then turned off. At the moment the drive tone is off, the pump tone $\omega_p$ is turned on by the other control pulse for a time $t_p$ to induce rabi oscillation between the two modes, just as shown in Figure 3 (a). Since the drive tone is off, the oscillation is induced by the residual phonons in the phonon cavity mode 2. The data is averaged over 2000 times.

**VI. Data processing of the phonon Rabi oscillation**

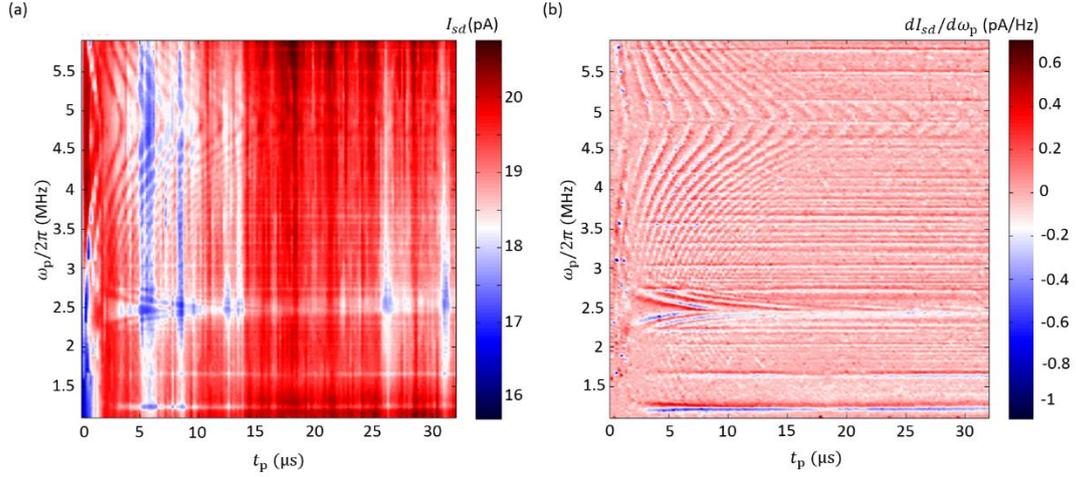

**Figure S5. Phonon Rabi oscillation measured by the transport current.** (a) The experimental data. The transport current $I_{sd}$ is recorded as scanning the pump frequency $\omega_p$ at different pump time $t_p$. The current noises change with the pump time. (b) The derivative of transport current $I_{sd}$ with respect to the pump frequency $\omega_p$ can effectively reduce the current noises and extract the phonon Rabi oscillation signal.

**VII. The explanation about the current is not sensitive to mode 1 at certain range of gate voltage**

We experimentally find that electrical signal of the in-plane mechanical mode 1 becomes weaker and weaker as the gate voltage is tuned away from the coulomb peak [Fig. 3 (c)]. There is also a slight but not obvious decrease of the signal of out-of-plane mechanical mode 2 at the same range of gate voltage. According to the mechanism of detecting the nanotube mechanical motion by quantum dot or SET in this supporting information I, we exclude the mutual term $\frac{d^2 I_{sd}^{DC}(V_g)}{dV_g^2}$ and $\frac{V_g}{C_g}$ in equation S3 because of the behavior of out-of-plane mechanical mode 2 and the only difference between the two modes is the term $\frac{\partial C_g}{\partial u_i}$ in equation S2 and S3. So we infer that $\frac{\partial C_g}{\partial u_i}$ is more sensitive to the change of gate voltage for the in-plane mechanical mode than that for the out-of-plane mechanical mode.

1. Deng, G. W.; Zhu, D.; Wang, X. H.; Zou, C. L.; Wang, J. T.; Li, H. O.; Cao, G.; Liu, D.; Li, Y.; Xiao, M.; Guo, G. C.; Jiang, K. L.; Dai, X. C.; Guo, G. P. *Nano letters* **2016,** 16, (9), 5456-62.
2. Lassagne, B.; Tarakanov, Y.; Kinaret, J.; Garcia-Sanchez, D.; Bachtold, A. *Science* **2009,** 325, (5944), 1107-10.
3. Steele, G. A.; Huttel, A. K.; Witkamp, B.; Poot, M.; Meerwaldt, H. B.; Kouwenhoven, L. P.; van der Zant, H. S. *Science* **2009,** 325, (5944), 1103-7.